\documentclass[11pt]{article}

\usepackage{amsthm,amsmath,amssymb}

\usepackage{supertabular}
\usepackage{hhline}
\usepackage{multicol}

\usepackage{graphicx}
\usepackage{hhline}
\usepackage{multicol}
\usepackage{array}
\usepackage{color}

\usepackage{placeins}
\usepackage{algorithmicx}
\usepackage[ruled]{algorithm}
\usepackage{algpseudocode}

\usepackage{hyperref}

\theoremstyle{plain}
\newtheorem{theorem}{Theorem}

\newtheorem{corollary}[theorem]{Corollary}

\theoremstyle{definition}
\newtheorem{definition}[theorem]{Definition}
\newtheorem{example}[theorem]{Example}

\newtheorem{notation}[theorem]{Notation}
\newtheorem{algorithmo}[theorem]{Algorithm}

\theoremstyle{remark}

\newtheorem{note}[theorem]{Note}

\title{\bf Regular Bipartite Graphs And Their Properties}

\author{Vivek S. Nittoor  \\
\small\tt vivek@nittoor.com
}

\begin{document}

\maketitle


\begin{abstract}
   We introduce a new notation for representing labeled regular bipartite graphs of arbitrary degree. Several enumeration problems for labeled and unlabeled regular bipartite graphs have been introduced. A general algorithm for enumerating all non-isomorphic $2$-regular bipartite graphs for a specified number of vertices has been described and a mathematical proof has been provided for its completeness. An abstraction of $m$ Symmetric Permutation Tree in order to visualize a labeled $r$-Regular Bipartite Graph with $2m$ vertices and enumerate its automorphism group has been introduced. An algorithm to generate the partition associated with two compatible permutations has been introduced. The relationship between Automorphism Group and permutation enumeration problem has been used to derive formulae for the number of compatible permutations corresponding to a specified partition. 
\end{abstract}

\label{chap_rbg}

The enumeration results for $2$-regular bipartite graphs and trivalent bipartite graphs developed in this chapter form the foundation for search for graphs with maximum girth.

A general algorithm for enumerating all non-isomorphic $2$-regular bipartite graphs with $2m$ vertices has been described and a mathematical proof has been provided for its completeness. An abstraction of $m$ Symmetric Permutation Tree, \nameref{notation_SPT} in order to visualize a labeled $r$-regular bipartite graphs with $2m$ vertices and enumerate its automorphism group has been introduced. An algorithm to generate the partition associated with two compatible permutations has been introduced. The relationship between Automorphism Group and permutation enumeration problem has been used to derive formulae for the number of compatible permutations corresponding to a specified partition.

We start with Introduction in Section \ref{sec_intro_chap_rbg}, introduce enumeration problems for regular bipartite graphs in Section \ref{sec_enu_prob_rbg}, and enumeration of non-isomorphic regular bipartite graphs has been explored in Section \ref{sec_non_iso_graph_enu}.

We introduce Symmetric Permutation Tree, \nameref{notation_SPT} and discuss its properties in Section \ref{sec_spt}, discuss cycles in Section \ref{sec_cycles}, Labeled Graph Enumeration in Section \ref{sec_labeled_graph_enumeration}, Permutation Enumeration Formulae in Section \ref{sec_permutation_enumeration_formulae}, Exhaustive enumeration of all leaf nodes of a Symmetric Permutation Tree in Section \ref{sec_towards_exhaus_enu}, and Conclusion in Section \ref{sec_rbg_chap_summ}.

\section{Introduction}
\label{sec_intro_chap_rbg}
We discuss preliminary enumeration results for $2$-regular bipartite graphs of $2m$ vertices, abstraction of the $m$ Symmetric Permutation Tree, interpretation of a labeled $r$-regular bipartite graphs of $2m$ vertices, permutations, partitions between permutations and finally the family of regular bipartite graphs $\Phi(\beta_{1}, \beta_{2}, \ldots, \beta_{r -1})$ has been introduced.

\begin{definition} Labeled $r$-regular bipartite graph \\
A Labeled $r$-regular bipartite graph with $2m$ vertices is a Bipartite Graph $G = (N, E)$ where the set of edges $E$ is a subset of the Cartesian product $E\subset \mathit{CN} \times  \mathit{VN}$ where $N= \mathit{CN}\ \cup\  \mathit{VN}$ and $ \mathit{CN}\ \cap\  \mathit{VN}\ =\emptyset$ such that the cardinality of the sets $ \mathit{CN}$ and $ \mathit{VN}$ are each $m$ and degree of each vertex is $r$, such that the $m$ vertices in the sets $ \mathit{CN}$ and $ \mathit{VN}$ have distinct labels.
\end{definition}

We use the following notation for labeling the vertices of a labeled regular bipartite graph.
Notation \ref{notation1_ver_rbg} where the two sets of vertices are labeled as follows, with $CN = \{\mathit{CN}_{1}, \mathit{CN}_{2}, \ldots,\mathit{CN}_{m}\}$ and 
$VN = \{\mathit{VN}_{1}, \mathit{VN}_{2}, \ldots, \mathit{VN}_{m}\}$.

\begin{notation} {Labels For Vertices of Labeled $r$-Regular Bipartite Graph with $2m$ Vertices} \\
\label{notation1_ver_rbg}
We refer to the two sets of $m$ vertices each as $CN = \{\mathit{CN}_{1}, \mathit{CN}_{2}, \ldots,\mathit{CN}_{m}\}$ and 
$VN = \{\mathit{VN}_{1}, \mathit{VN}_{2}, \ldots, \mathit{VN}_{m}\}$.
\end{notation}

Bi-adjacency matrix is a very important and yet natural way to specify a labeled bipartite graph. We define bi-adjacency matrix for a labeled bipartite graph in Definition \ref{definition_bi_adjacency_matrix}.

\begin{definition} {Bi-adjacency Matrix For A Labeled Bipartite Graph} \\
\label{definition_bi_adjacency_matrix}
Let the labeled regular bipartite graph be $(N, E)$ with set of edges $E$ and set of vertices $N$, where $N = CN\ \cup\ VN$ and $CN\ \cap\ VN = \emptyset$ such that $CN =\{CN_{1}, CN_{2},\ldots, CN_{m}\}$ and $VN =\{VN_{1}, VN_{2},\ldots, VN_{m}\}$ are set of vertices and the set of edges $E$ is the subset of the cartesian product of $CN$ and $VN$ given by $CN \times VN$.
The bi-adjacency matrix of a regular bipartite graph with order $2m$ is a $m \times m$ matrix with element $c_{i,j}=1$ if and only if $(CN_{i}, VN_{j})\in E$ and $c_{i,j}=0$ if and only if $(CN_{i}, VN_{j})\notin E$.
\end{definition}

A $r$-regular bipartite graph with $2m$ vertices can be represented by its bi-adjacency matrix, a $m \times m$ square matrix with $r$ non-zero elements in each of its rows and columns. There exists a one-one correspondence between a regular bipartite graph and its equivalent bi-adjacency matrix representation. We divide the two sets of vertices in a regular bipartite graph, 
each of which are $m$ in number.

In general, a $r$-regular bipartite graph with $2m$ vertices is labeled when the two disjoint set of $m$ vertices are labeled with unique labels. We introduce an additional constraint in the labeling process as follows.

In the bi-adjacency matrix representation of a labeled $r$-regular bipartite graph with $2m$ vertices, element $(i, j)$ takes the value $1$ if vertex $i$ where $1\le i \le m$ from set $CN$ is connected to vertex $j$ where $1\le j \le m$ from set $VN$ and is $0$ otherwise.

\begin{definition} Isomorphic Graphs \\
\label{def_iso_graphs}
Two graphs $G_{1} = (N_{1}, E_{1})$ and $G_{2}= (N_{2}, E_{2})$ are isomorphic if there exists an one-one onto map between vertex sets $N_{1}$ and $N_{2}$ that preserve adjacency. 
\end{definition}

Definition \ref{def_iso_bi_graphs} for isomorphic regular bipartite graphs has been derived from Definition \ref{def_iso_graphs} for Isomorphic Graphs.

\begin{definition}{Isomorphic Regular Bipartite Graphs} \\
\label{def_iso_bi_graphs}
Two labeled regular bipartite graphs with bi-adjacency matrices $A$ and $B$ are isomorphic to each other if there exists a set of row and
column exchange operations that transform $A$ into $B$.
\end{definition}

\subsubsection{Self-evident Facts About Isomorphic Regular Bipartite Graphs}
If two labeled regular bipartite graphs with bi-adjacency matrices $A$ and $B$ are isomorphic to each other, then the following statements are true.
\begin{enumerate}
\item The two labeled regular bipartite graphs with bi-adjacency matrices $A$ and $B$ have the same regular node degree.
\item The two labeled regular bipartite graphs with bi-adjacency matrices $A$ and $B$ have the same number of vertices.
\end{enumerate}

\begin{definition} {Automorphism group of a graph} \\
An automorphism of a graph $G=(V,E)$ is a permutation $\sigma $ of the vertex set $V$, such that for any edge $e=(u,v)$; we also have $\sigma (e)=(\sigma u,\sigma v)$ as an edge of the graph $G$. Set of automorphisms of a graph form a group and this group is referred to as the automorphism group of the graph.
\end{definition}

The set of all labeled $r$-regular bipartite graphs with $2m$ vertices that are isomorphic to a given labeled $r$-regular bipartite graph with $2m$ vertices constitute its Automorphism Group with the group operation defined as the isomorphism between two labeled $r$-regular bipartite graph with $2m$ vertices.

\begin{definition}{Non-isomorphic $r$-Regular Bipartite Graphs with $2m$ Vertices} \\
Two labeled $r$-regular bipartite graphs with $2m$ vertices with bi-adjacency matrices $A$ and $B$ are non-isomorphic if there does not exist any row or column
exchanges which can transform $A$ into $B$.
\end{definition}

Example \ref{example_Iso_gra} shows bi-adjacency matrices of isomorphic graphs and Example \ref{example_Non_Iso_gra} shows bi-adjacency matrices of non-isomorphic graphs.

\begin{example} \textbf{Example Of Isomorphic Graphs} \\
\label{example_Iso_gra}
\begin{equation*}
\left[\begin{matrix}0&0&0&0&1&1\\1&0&0&0&0&1\\1&1&0&0&0&0\\0&0&1&1&0&0\\0&0&0&1&1&0\\0&1&1&0&0&0\end{matrix}\right]\equiv
\left[\begin{matrix}1&0&0&0&0&1\\1&1&0&0&0&0\\0&1&1&0&0&0\\0&0&1&1&0&0\\0&0&0&1&1&0\\0&0&0&0&1&1\end{matrix}\right]
\end{equation*}
\end{example}

\begin{example} \textbf{Example Of Non-Isomorphic Graphs} \\
\label{example_Non_Iso_gra}
\begin{equation*}
\left[\begin{matrix}0&0&0&0&1&1\\1&0&0&0&0&1\\1&1&0&0&0&0\\0&0&1&1&0&0\\0&0&0&1&1&0\\0&1&1&0&0&0\end{matrix}\right]\not\equiv
\left[\begin{matrix}1&0&1&0&0&0\\1&1&0&0&0&0\\0&1&1&0&0&0\\0&0&0&1&0&1\\0&0&0&1&1&0\\0&0&0&0&1&1\end{matrix}\right]
\end{equation*}
\end{example}
All labeled $r$-regular bipartite graphs with $2m$ vertices that are isomorphic to each other form an equivalence class that we shall later define as the automorphism group of a labeled $r$-regular bipartite graph with $2m$ vertices. We usually refer to a representative member from the equivalence class of $r$-regular bipartite graphs with $2m$ vertices that are isomorphic to each other as the canonical form.

\section{Enumeration Problems For Regular Bipartite Graphs}
\label{sec_enu_prob_rbg}

We introduce two enumeration problems for regular bipartite graphs as follows.
\begin{enumerate}
\item Non-Isomorphic $r$-Regular Bipartite Graph Enumeration Problem.
\item Non-Isomorphic $r$-Regular Bipartite Graph Enumeration Problem.
\end{enumerate}

\begin{definition} {Labeled $r$-Regular Bipartite Graph Enumeration Problem} \\
Labeled Enumeration of $r$-regular bipartite graphs with $2m$ vertices refers to enumeration of all distinct labeled $r$-regular bipartite graphs with $2m$ vertices.
\end{definition}

A subset of this problem would be to enumerate all labeled $r$-regular bipartite graphs with $2m$ vertices that are isomorphic to a given labeled $r$-regular bipartite graph with $2m$ vertices.
\begin{definition} {Non-Isomorphic $r$-Regular Bipartite Graph Enumeration Problem}\\
Non-Isomorphic Enumeration of $r$-regular bipartite graphs with $2m$ vertices is Enumeration of all distinct non-isomorphic $r$-regular bipartite graphs with $2m$ vertices.
\end{definition}

\begin{definition} Enumeration function $E(m,r)$ for a $r$-regular bipartite graph with $2m$ vertices \\
Let $E(m,r)$ where $r\le m$ be a $E:N^{2}\to N\text{~}\text{~}$ function that represents enumerations of
number of distinct non-isomorphic $r$-regular bipartite graphs with $2m$ vertices. 
\end{definition}

\subsection{List of partitions}
Let $P_{2}(m)$ be a set of partitions of $m$ using natural numbers that are greater or equal to $2$. 

\begin{definition}{$P_{2}(m)$} \\
\label{definition_p2m}
$P_{2}(m)$ is defined as a set of partitions of $m$ that consist of natural numbers that are greater or equal to $2$. 
\end{definition}

\begin{definition}{Partition Component For a Partition $\beta \in P_{2}(m)$} \\
\label{definition_parcomp}
If $\beta \in P_{2}(m)$ refers to $\sum_{j=1}^{y} q_{j}=m$, then each ${q_{j}}$ for $ 1\le j \le y $ is referred to as a partition component of $\beta$. 
\end{definition}

Let us consider the partitions of $m$ that consist of numbers greater than or equal to $2$ in Example \ref{ex_p2_m}.

\begin{example} \textbf{$P_{2}(m)$ for various values of $m$} \\
\label{ex_p2_m}
\begin{center}
\begin{tabular}{|r|r|}
\hline
\multicolumn{1}{|c|}{$
m=4
$
} & \multicolumn{1}{c|}{$
\{(2,2);(4)\}
$
}\\\hline
\multicolumn{1}{|c|}{$
m=5
$
} & \multicolumn{1}{c|}{$
\{(3,2);(5)\}
$
}\\\hline
\multicolumn{1}{|c|}{$
m=6
$
} & \multicolumn{1}{c|}{$
\{(2,2,2);(3,3);(4,2);(6)\}
$
}\\\hline
\multicolumn{1}{|c|}{$
m=7
$
} & \multicolumn{1}{c|}{$
\{(3,2,2);(4,3);(5,2);(7)\}
$
}\\\hline
\multicolumn{1}{|c|}{$
m=8
$
} & \multicolumn{1}{c|}{$
\begin{gathered}\{(2,2,2,2);(4,2,2);(6,2);(4,4);\\(5,3);(3,3,2);(8)\}\end{gathered}
$
}\\\hline
\multicolumn{1}{|c|}{$
m=9
$
} & \multicolumn{1}{c|}{$
\begin{gathered}\{(3,2,2,2);(4,3,2);(6,3);(5,4);\\(6,3);(7,2);(3,3,3);(9)\}\end{gathered}
$
}\\\hline
\end{tabular}
\end{center}
\end{example}

\section{Non-Isomorphic Graph Enumeration}
\label{sec_non_iso_graph_enu}

A general algorithm for enumerating all non-isomorphic $2$-regular bipartite graphs with $2m$ vertices is described in this section. It is clear that we have only one non-isomorphic graph for $r=1$, the canonical form of which could be represented as $I_{m}$, which is a $m \times m$ identity matrix. It is clear that we have $m!$ labeled $1$-regular bipartite graphs with $2m$ vertices that are isomorphic to each other by considering all the $m!$ permutations on labels.

\subsubsection{Mapping Between Partitions And Regular Bipartite Graphs}

Theorem \ref{thm_m_2_btu} solves the enumeration problem of non-isomorphic $2$-regular bipartite graphs, and is an important step towards solving the enumerating $3$-regular bipartite graphs. The enumeration algorithm \ref{algo_m_2_btu} for enumerating all non-isomorphic $2$-regular bipartite graphs is based upon Theorem \ref{thm_m_2_btu}.

\begin{theorem}
\label{thm_m_2_btu}
Each element $\beta \in P_{2}(m)$ corresponds to a non-isomorphic $2$-regular bipartite graph with $2m$ vertices. The number of non-isomorphic $2$-regular bipartite graphs with $2m$ vertices is precisely the number of elements in $P_{2}(m)$, i.e., $E(m,2)=p(m,2)\ \forall\ m>2,m\in \mathbb{N}$.
\end{theorem}

\begin{proof}
Let us consider a labeled $2$-regular bipartite graph with $2m$ vertices having two distinct set of nodes $\mathit{CN}_{1},\mathit{CN}_{2},\ldots ,\mathit{CN}_{m}$ and $\mathit{VN}_{1},\mathit{VN}_{2},\ldots, \mathit{VN}_{m}$. We grow this labeled regular bipartite graph by in steps by initially creating a labeled $1$-regular bipartite graph with $2m$ vertices, and then a labeled $2$-regular bipartite graph with $2m$ vertices.

For $r=1$, without loss of generality, we connect node $\mathit{CN}_{i}$ to node $\mathit{VN}_{i}$  $\ \forall\ 1\le i\le m$.
We have only one non-isomorphic graph for $r=1$ since we can permute the vertices $\mathit{VN}_{1},\mathit{VN}_{2},\ldots, \mathit{VN}_{m}$ among themselves before any connection is made, and if 
$\mathit{CN}_{i}$ is connected to $\mathit{VN}_{j}$, we permute $\mathit{VN}_{i}$ and $\mathit{VN}_{j};j>i$, and hence the first
edge connection is made between each $\mathit{CN}_{i}$ and $\mathit{VN}_{i}$ $\ \forall\ 1\le i\le m$. Hence, without loss of
generality, for $r=1$, the first edge connection is made between each $\mathit{CN}_{i}$ and $\mathit{VN}_{i}$ $\ \forall\ 1\le i\le m$.
This could be expressed as follows.\\ 
for( $i=1$; $i\le m$; $i$++)\{ \\
{\ \ \ Connect $\mathit{CN}_{i}$ with $\mathit{VN}_{i}$; }\\
\} \\
For $r=2$, starting from $\mathit{CN}_{1}$, we connect the second edge from $\mathit{CN}_{1}$ to $\mathit{VN}_{i(1)}$ where 
$1<i(1)\le m$ ; and then $\mathit{CN}_{i(1)}$ to an arbitrary VN node $\mathit{VN}_{i(2)};i(2)\neq i(1)\neq 1$. Connect, 
$\mathit{CN}_{i(2)}$ to $\mathit{VN}_{i(3)}$, and so on until $\mathit{CN}_{i(k)}$ to $\mathit{VN}_{1}$, until $\mathit{VN}_{1}$ 
is reached for some positive integer $k;1\le k\le m$.

If all the $\mathit{CN}$ and $\mathit{VN}$ vertices do not have two edges, let $\mathit{CN}_{i(1,0)}$ be the $\mathit{CN}$ node which
does not have two edges such that $i(1,0)$ is the minimum value of indexes for all vertices that do not yet have two edges.

Starting from $\mathit{CN}_{i(1,0)}$, we connect the second edge to an arbitrary VN node which has only one edge, 
$\mathit{VN}_{i(1,1)};i(1,1)\neq i(1,0)$. Similarly, we connect $\mathit{CN}_{i(1,1)}$ to $\mathit{VN}_{i(1,2)};i(1,2)\neq i(1,1)\neq
i(1,0)$, and so on until $\mathit{VN}_{i(1,0)}$ is reached. We continue the above process until all the $\mathit{CN}$ and 
$\mathit{VN}$ vertices have two edges each and we have a $2$-regular bipartite graph with $2m$ vertices.
We can establish a one-one onto map corresponding with this structure and a partition of $m$ that consist of numbers that are greater than or equal
to $2$.
Thus, we have established a mapping between an arbitrary unlabeled $2$-regular bipartite graph with $2m$ vertices and partitions of $m$ that consist of
numbers that are greater than or equal to $2$.

We now consider the mapping between partitions and enumerations for $2$-regular bipartite graphs with $2m$ vertices.

If we consider a partition $(p_{1},p_{2},\ldots, p_{y})\in P_{2}(m)$ where $\sum _{i=1}^{y}p_{i}=m$ and $p_{i}\ge 2$ for $1\le i\le y\in \mathbb{N}$, now consider a $2$-regular bipartite graph with $2m$ vertices.

Without loss of generality, we assume that the first edge are connected
from $\mathit{CN}_{j}$ to $\mathit{VN}_{j}$ for $1\le j\le m$.

For the second edge for each $\mathit{VN}$ and $\mathit{CN}$ node,

for each $1\le j<p_{1}$ we connect $\mathit{CN}_{j}$ to $\mathit{VN}_{j+1}$ and then connect $\mathit{VN}_{1}$ to 
$\mathit{CN}_{p_{1}}$. 
For each $p_{1}+p_{2}+\ldots +p_{i}\le j<p_{1}+p_{2}+\ldots+p_{i}+p_{i+1}$, we connect $\mathit{CN}_{j}$ to $\mathit{VN}_{j+1}$ and then
Connect $\mathit{VN}_{p_{1}+p_{2}+\ldots +p_{i}}$ to 
$\mathit{CN}_{p_{1}+p_{2}+\ldots +p_{i}+p_{i+1}}$
For each $p_{1}+p_{2}+\ldots +p_{y-1}\le j<p_{1}+p_{2}+\ldots +p_{y}=m$
we connect $\mathit{CN}_{j}$ to $\mathit{VN}_{j+1}$ and then
connect $\mathit{VN}_{p_{1}+p_{2}+\ldots +p_{y-1}}$ to 
$\mathit{CN}_{m}$.

Thus, any partition of $m$, that consists of natural numbers that are
greater or equal to $2$ could be mapped to a $2$-regular bipartite graph. Thus, any possible structure of the 
$2$-regular bipartite graph with $2m$ vertices could be mapped to a partition of $m$, that consists of natural numbers that are greater or equal to $2$.
\end{proof}

\begin{corollary}
$E(m,2)=p(m,2)\ \forall \ m>2,m \in \mathbb{N}$ where $p(m,r)$ 
represents the number of partitions of $m$ using natural numbers that
are greater or equal to $r$.
\end{corollary}

\begin{corollary}
The labeled $2$-regular bipartite graph with $2m$ vertices represented by compatible permutations $p_{1},p_{2}\in S_{m};p_{2}\notin C(p_{1})$ is isomorphic to $\Psi (\beta )$ for some $\beta \in P_{2}(m)$.
\end{corollary}

\begin{definition} {Canonical forms for $2$-Regular Bipartite Graphs with $2m$ vertices} \\
Non-isomorphic forms of $2$-regular bipartite graphs with $2m$ vertices correspond to
each of the partitions of $m$ that consists of natural numbers that are greater or equal to $2$.
\end{definition}

The canonical forms for $r=2$ correspond to the matrices generated by algorithm \ref{algo_canonical}. 

\begin{algorithmo} {\label{algo_canonical}} \textbf{Construction Of Canonical Forms of 2-Regular Bipartite Graphs} \\
\textbf{Input}: $m \in \mathbb{N}$ and $\beta \in P_{2}(m)$.\\
\textbf{Output}: $\Psi(\beta)$, $2$-regular bipartite graph with $2m$ vertices. \\
\textbf{Method}: The procedure $ConstructDegree2\_\Psi(m)$ as shown in Figure \ref{func_const_deg2_psi}, enumerates all elements of the set $P_{2}(m)$ and then calls
procedure $Construct(\beta)$ as shown in Figure \ref{func_const_beta} to construct a $2$-regular bipartite graph with $2m$ vertices with $\Psi(\beta)$ for each $\beta \in P_{2}(m)$.
\bigskip
\begin{figure}[htbp]
\centering
\vspace{-0.2cm}%
\alglanguage{pseudocode}
\begin{algorithmic}[1]
\Procedure{$ConstructDegree2\_\Psi$}{$m$}
	\State Enumerate all elements of the set $P_{2}(m)$
\For{each $\beta \in P_{2}(m)$}
	\State Construct$(\beta)$
\EndFor
\EndProcedure
\Statex
\end{algorithmic}
  \vspace{-0.4cm}%
\caption{Canonical Form of $2$-Regular Bipartite Graphs with $2m$ vertices}
\label{func_const_deg2_psi}
\end{figure}

\bigskip
\begin{figure}[htbp]
\centering
\vspace{-0.2cm}%
\alglanguage{pseudocode}
\begin{algorithmic}[1]
\Procedure{$Construct$}{$\beta$} 
	\State Enumerate all partition components of $\beta \in P_{2}(m)$ in ascending order
	\State Initialize $d$ to the number of partition components of $\beta$
	\State $s = 0$ 
	\State $t = 0$ 
\For{$i= 0 \to  m - 1,  i++$} 
\State  Connect $\mathit{CN}_{i}$ with $\mathit{CN}_{i}$
\EndFor
\For{$j= 0 \to  d - 1,  j++$} 
	\State  Initialize $u$ to the next partition components of $\beta$
\For{$z= 0 \to  u - 1,  j++$} 
	\State  $l = (z + 1) \% u + s$
	\State  Connect $\mathit{CN}_{t}$ with $\mathit{CN}_{t}$
	\State  $t++$
\EndFor
	\State $s = s + u$
\EndFor
\EndProcedure
\Statex
\end{algorithmic}
  \vspace{-0.4cm}%
\caption{Construction of $2$-Regular Bipartite Graph $\Psi(\beta)$ for $\beta \in P_{2}(m)$}
\label{func_const_beta}
\end{figure}

\end{algorithmo}

\begin{notation} {Notation $\Psi$ for the Canonical form Of A $2$-Regular Bipartite Graph with $2m$ vertices generated by $\beta \in P_{2}(m)$} \\
\label{notation_psi}
We denote the $2$-regular bipartite graph with $2m$ vertices generated by $\beta \in P_{2}(m)$ as $\Psi(\beta)$. $\Psi$ is thus a function that maps each element from $P_{2}(m)$ to a $2$-regular bipartite graph with $2m$ vertices, each of which would be a Canonical Form in its Eqivalence Class.\\
$\Psi :P_{2}(m)\to$ Set Of $2$-Regular Bipartite Graphs with $2m$ vertices.
\end{notation}

\begin{theorem}
The number of elements in the set $P_{2}(m)$ is given by $p(m,2)=p(m) - p(m-1)$ where $p(m)$ is the number of unrestricted partitions of $m\in \mathbb{N}$.
\end{theorem}
\begin{proof} 
If $P(m)$ is the set of all unrestricted partitions of $m\in \mathbb{N}$, we can establish a one-one onto map between the set $P(m-1)$ and subset $A(m)\subset P(m)$ that contains partitions of 
$m$ with at least one $1$ in it since all partitions of $m-1$ with $1$ appended are partitions of $m$ with at least one $1$. Hence, the
set $P_{2}(m)$ is generated when $A(m)$ is removed from $P(m)$ and therefore we obtain the equation $E(m,2)=p(m, 2)=p(m) - p(m-1)$.
\end{proof}
The earliest equation for $p(m)$ was obtained by Ramanujan and Hardy in $1918$ and has been described in \cite{27}.

\begin{example} \textbf{Sample enumeration all non-isomorphic $2$-Regular Bipartite Graphs with $12$ Vertices}\\
For example, $m=6$, we obtain $P_{2}(6)=\{(6),(4,2),(3,3),(2,2,2)\}$
and therefore all non-isomorphic $2$-regular bipartite graphs with $12$ vertices can be enumerated as shown in Figure \ref{fig_all_non_iso_6}.

\begin{figure}[htbp]
\centering
\begin{eqnarray*}
\nonumber
\Psi((6))=\left[\begin{matrix}1&0&0&0&0&1\\1&1&0&0&0&0\\0&1&1&0&0&0\\0&0&1&1&0&0\\0&0&0&1&1&0\\0&0&0&0&1&1\end{matrix}\right] \\
\nonumber
\Psi((2,2,2))=\left[\begin{matrix}1&1&0&0&0&0\\1&1&0&0&0&0\\0&0&1&1&0&0\\0&0&1&1&0&0\\0&0&0&0&1&1\\0&0&0&0&1&1\end{matrix}\right] \\
\nonumber
\Psi((4,2))=\left[\begin{matrix}1&0&0&1&0&0\\1&1&0&0&0&0\\0&1&1&0&0&0\\0&0&1&1&0&0\\0&0&0&0&1&1\\0&0&0&0&1&1\end{matrix}\right] \\
\nonumber
\Psi((3,3))=\left[\begin{matrix}1&0&1&0&0&0\\1&1&0&0&0&0\\0&1&1&0&0&0\\0&0&0&1&0&1\\0&0&0&1&1&0\\0&0&0&0&1&1\end{matrix}\right] \\
\end{eqnarray*}
\caption{All Non-Isomorphic $2$-Regular Bipartite Graphs With $12$ Vertices}
\label{fig_all_non_iso_6}
\end{figure}
\end{example}

\begin{algorithmo} \textbf{{Enumeration Algorithm for Degree $2$}} \\
\label{algo_m_2_btu}
Algorithm to enumerate all non-isomorphic $2$-regular bipartite graphs with $2m$ vertices is as follows.\\
\textbf{Input}: $m \in \mathbb{N}$.\\
\textbf{Output}: All all non-isomorphic $2$-regular bipartite graphs with $2m$ vertices. \\
\textbf{Method}: The procedure $EnumerateAllNonIsomorphicDegree2(m)$ as shown in Figure \ref{func_enu_all_non_iso_deg2}
\begin{enumerate}
\item Enumerate all elements of the set $P_{2}(m)$.
\item For each instance of $\beta \in P_{2}(m)$, construct a $2$-regular bipartite graph with $2m$ vertices with $\Psi(\beta)$ as per Algorithm \ref{algo_canonical}.
\end{enumerate}

\bigskip
\begin{figure}[htbp]
\centering
\vspace{-0.2cm}%
\alglanguage{pseudocode}
\begin{algorithmic}[1]
\Procedure{$EnumerateAllNonIsomorphicDegree2$}{$m$} 
	\State Enumerate all elements of the set $P_{2}(m)$
\For{each $\beta \in P_{2}(m)$}
	\State Construct $\Psi(\beta)$
\EndFor
\EndProcedure
\Statex
\end{algorithmic}
  \vspace{-0.4cm}%
\caption{Enumeration Of All Non-Isomorphic $2$-Regular Bipartite Graphs}
\label{func_enu_all_non_iso_deg2}
\end{figure}

\end{algorithmo}

\subsection{Rank Of A Regular Bipartite Graph}

We consider the rank of the bi-adjacency matrix of a $r$-regular bipartite graph with $2m$ vertices in $\mathit{GF}(2)$.
The well known set $\mathit{GF}(2)$, is the Galois field of two elements that consists of $0$ and $1$ with operations of modular addition and modular multiplication.

\begin{definition} Rank of a $r$-Regular Bipartite Graph with $2m$ Vertices\\
The rank of a $r$-regular bipartite graph with $2m$ vertices is defined as the rank of its bi-adjacency matrix in $\mathit{GF}(2)$, the Galois field of two elements that consists of $0$ and $1$ with operations of modular addition and modular multiplication.
\end{definition}

\subsection{2-Regular Bipartite Graph Rank Theorem} 
\begin{theorem}
If a $2$-regular bipartite graph with $2m$ vertices is constructed with partition $\beta \in P_{2}(m)$ that refers to $\sum _{i=1}^{k}q_{i}=m$, 
then its rank is $m-k$ in $\mathit{GF}(2)$.
\end{theorem}

\begin{proof}
A partition $\sum _{i=1}^{k}q_{i}=m$ will correspond to $k$ components in the $2$-regular bipartite graph with $2m$ vertices, and for each component, we
have the first row as sum of all other rows in $\mathit{GF}(2)$ as per definition of $\Psi$. Hence rank of each
component is $q_{i}-1$, Hence, rank of the $2$-regular bipartite graph with $2m$ vertices is $\sum _{i=1}^{k}(q_{i}-1)=m-k$.
\end{proof}

\begin{theorem}
A $r$-regular bipartite graph with $2m$ vertices is not full rank in $\mathit{GF}(2)$ if $r$ is an even positive integer.
\end{theorem}
\begin{proof}
One can verify that the sum of all rows of the $r$-regular bipartite graph with $2m$ vertices is $0$ in $\mathit{GF}(2)$ if $r$ is even. Hence, a $r$-regular bipartite graph with $2m$ vertices is not full rank if $r$ is even.
\end{proof}

\subsection {Properties Of E(m, r)}
\begin{theorem}
\label{thm_e_m_r_m_minus_r}
The Enumerations of $r$-regular bipartite graphs with $2m$ vertices correspond to enumerations of partitions in the following manner.
\begin{equation}
E(m, r)=E(m, m - r)\  \forall\ r ;0\le r\le m
\end{equation}
\end{theorem}

\begin{proof}
We can establish a one-one onto map between the set of all non-isomorphic $r$-regular bipartite graphs with $2m$ vertices and the set of all
non-isomorphic $(m- r)$-regular bipartite graphs with $2m$ vertices for $0\le r\le m$ in the following manner. Let us assume that $E(m,r)=z$. If
$\{A_{1},A_{2},\ldots, A_{z}\}$ are the canonical forms for a non-isomorphic $r$-regular bipartite graphs with $2m$ vertices, then we obtain bi-adjacency matrices for
$m-r$-regular bipartite graphs with $2m$ vertices $\{B_{1},B_{2},\ldots, B_{z}\}$ by mapping all $0$s in each $A_{i}$ to $1$s and mapping all $1$s in each $A_{i}$ to $0$s. 
In the set $\{B_{1},B_{2},\ldots, B_{z}\}$, we observe that $B_{i}\not\equiv B_{j}$ for $i\neq j;1\le i\le z;i\le j\le z$. If possible, let there exist a
$m-r$-regular bipartite graph with $2m$ vertices, matrix $C$ such that $C$ is not isomorphic to any element in $\{B_{1},B_{2},\ldots,B_{z}\}$. 
Let us map $(m-r)$-regular bipartite graph with $2m$ vertices, bi-adjacency matrix $C$ to $r$-regular bipartite graph with $2m$ vertices, matrix $D$ by mapping all
$0$s in $C$ to $1$s and all $1$ in each $C$ to $0$s. Since $D$ is a $r$-regular bipartite graph with $2m$ vertices, it must be isomorphic to some element
$A_{j}$ of $\{A_{1},A_{2},\ldots, A_{z}\}$. Hence $C$ is isomorphic to $B_{j}$. Hence, there cannot exist a $(m-r)$-regular bipartite graph with $2m$ vertices that is not isomorphic to any element in $\{B_{1},B_{2},\ldots, B_{z}\}$ and we have already shown that no two elements in $\{B_{1},B_{2},\ldots, B_{z}\}$ are isomorphic to each other. Hence $E(m,m - r)=z$.
Hence, $E(m,r)=E(m,m-r)\ \forall\ r;0\le r\le m$.
\end{proof}

\subsection{Number of Non-Isomorphic Graphs Checked Manually}

We have manually worked out all possible non-isomorphic graphs for regular bipartite graphs for values of $m$ between $4$ and $6$. Number of non-isomorphic $r$-regular bipartite graphs with $2m$ vertices are shown in Table \ref{table_E_m4_r_manual_chk} for $m = 4$, Table \ref{table_E_m5_r_manual_chk} for $m = 5$ and Table \ref{table_E_m6_r_manual_chk} for $m = 6$. Determining the values of $E(m, r)$ for values of $m$ greater than $6$ for $r \ge 3$ is not feasible using manual methods. The values of $E(m, r)$ obtained for $r = 1$ have found to be consistent with $E(m, 1) = 1$ and the values of $E(m, r)$ obtained for $r = 2$ have found to be consistent with $E(m, 2) = p(m, 2)$ as per Theorem \ref{thm_m_2_btu}. The values for $E(6, 2) = E(6, 4) = 4$ in Table \ref{table_E_m6_r_manual_chk} are consistent with $E(m, r) = E(m, m - r)$ for $m < r$ as per Theorem \ref{thm_e_m_r_m_minus_r}. 

\begin{table}
\centering
\caption{Number of Non-Isomorphic Graphs Checked Manually for $m = 4$}
\begin{tabular}{lllllll}
\hline\noalign{\smallskip}
 $m$ &
 $r$ &
$E(m, r)$  \\
\noalign{\smallskip}
\hline
\noalign{\smallskip} 
\label{table_E_m4_r_manual_chk}
4 & 1 & 1\\
4 & 2 & 2 \\
& 
 \\
\hline
\end{tabular}
\end{table}

\begin{table}
\centering
\caption{Number of Non-Isomorphic Graphs Checked Manually for $m = 5$}
\begin{tabular}{lllllll}
\hline\noalign{\smallskip}
 $m$ &
 $r$ &
$E(m, r)$  \\
\noalign{\smallskip}
\hline
\noalign{\smallskip} 
\label{table_E_m5_r_manual_chk}
5 & 1 & 1 \\
5 & 2 & 2 \\
& 
 \\
\hline
\end{tabular}
\end{table}

\begin{table}
\centering
\caption{Number of Non-Isomorphic Graphs Checked Manually for $m = 6$}
\begin{tabular}{lllllll}
\hline\noalign{\smallskip}
 $m$ &
 $r$ &
$E(m, r)$  \\
\noalign{\smallskip}
\hline
\noalign{\smallskip} 
\label{table_E_m6_r_manual_chk}
6 & 1 & 1 \\
6 & 2 & 4 \\
6 & 3 & 7 \\
6 & 4 & 4 \\
& 
 \\
\hline
\end{tabular}
\end{table}

\begin{note}{\textbf{Summarizing Properties of Enumeration Function $E(m, r)$}} \\
\begin{eqnarray}
E(m,r)&=&E(m, m - r) \\
E(m,1)&=&E(m, m - 1)=1 . \\
E(m,2)&=&p(m, 2) = p(m) - p(m -1) \\
E(m,2)&=&E(m,m - 2)=p(m,2) \\
E(m,r+1)&>&E(m,r) \ \mathit{for}\ 2<r<m/2 \\
E(m,r+1)&<&E(m,r) \ \mathit{for}\ m/2<r<m
\end{eqnarray}
\end{note}

\section{Symmetric Permutation Tree And Its Properties}
\label{sec_spt}

We introduce an enumeration tree called Symmetric Permutation Tree which is defined in Definition \ref{definition_spt}. The primary motivation of introducing this enumeration tree is the Symmetric Group on $m$ elements that has order $\{m!\}$. As shown in Table \ref{table_nodes_depths_spt}, a $m$ Symmetric Permutation Tree has $\{m!\}$ leaf nodes. The Symmetric Permutation Tree is a very important step towards solving the enumeration problem for $3$-regular bipartite graphs, and thus forms a part of the foundation of our research.

\begin{notation}
\subsubsection{SPT($m$)}
\label{notation_SPT}
We use the notation \nameref{notation_SPT} to refer to a $m$ Symmetric permutation tree in Definition \ref{definition_spt}.
\end{notation}

\begin{definition}{$m$ Symmetric permutation tree} \\
\label{definition_spt}
{A $m$ Symmetric permutation tree \nameref{notation_SPT} is defined as a labeled tree with the following properties.}
\begin{enumerate}
\item {
 \nameref{notation_SPT} has a single root node labeled $0$.}
\item {
 \nameref{notation_SPT} has $m$ nodes at depth $1$ from the root
node.}
\item {
 \nameref{notation_SPT} has nodes at depths ranging from $1$ to $m$,
with each node having a labels chosen from $\{1,2,\ldots, m\}$. The
root node $0$ has $m$ successor nodes. Each node at depth $1$ has
 $m -1$ successor nodes at depth $2$. Each node at depth $i$ has 
$m - i+1$ successor nodes at depth $i+1$. Each node at depth 
$m - 1$ has $1$ successor node at depth $m$.}

\item {
No successor node in \nameref{notation_SPT} has the same node label as
any of its ancestor nodes.}
\item {
No two successor nodes that share a common parent node have the same
label. }
\item {
The sequence of nodes in the path traversal from the node at depth $1$ 
to the leaf node at depth $m$ in \nameref{notation_SPT} represents
the permutation represented by the leaf node.}
\item {
 \nameref{notation_SPT} has $m!$ leaf nodes each of which represent an
element of the symmetric group of degree $m$ denoted by $S_{m}$.}
\end{enumerate}
\end{definition}

\begin{table}
\centering
\caption{Nodes at different depths of a $m$ Symmetric Permutation Tree}
\begin{tabular}{llllll}
\hline\noalign{\smallskip}
Node Depth & Nodes at specified depth & Successors per node at specified depth \\
\noalign{\smallskip}
\hline
\noalign{\smallskip}
\label{table_nodes_depths_spt}
$1$ & $m$ & $m - 1$ \\
$2$ & $m\ast (m -1)$ & $m - 2$ \\
$i$ & $\prod _{l=0}^{i-1}(m-l)$ & $m-i+1 $ \\
$m - 1$ & $(m - 1)!$ & $1$\\
$m$ & $m!$ & 0 \\
\hline
\end{tabular}
\end{table}

\begin{theorem}
There exists an one-one onto map between the set of permutations represented by the leaf nodes of a symmetric permutation tree
\nameref{notation_SPT} and elements of $S_{m}$, the Symmetric group of degree $m$.
\end{theorem}

\begin{proof} 
\nameref{notation_SPT} and $S_{m}$ have $m!$ elements
each, and and since the sequence of nodes in the path traversal from
the node at depth $1$ to the leaf node at depth $m$ in 
\nameref{notation_SPT} represents the permutation represented by the
leaf node, we can establish a one-one onto map between \nameref{notation_SPT} and $S_{m}$.
\end{proof} 

\begin{theorem}
Given any node other than the root node of a $m$ symmetric permutation tree \nameref{notation_SPT}, the set of all of its descendant nodes and the set of all sibling nodes for each node from depths $1$ to $m$ have the same number nodes and both sets have the same set of distinct labels.
\end{theorem}

\begin{proof} 
The set of all sibling nodes at any depth in a complete $m$ symmetric permutation tree \nameref{notation_SPT} contains all nodes from the set 
$\{1,2,\ldots, m\}$ except node labels of ancestor nodes. The descendant nodes at any depth of a $m$ symmetric permutation tree \nameref{notation_SPT}
contains all nodes from the set $\{1,2,\ldots, m\}$ except node labels of ancestor nodes. Hence, the sibling nodes and the descendant nodes for each node in a complete $m$ symmetric permutation tree contains the same number of elements with precisely the same labels.
\end{proof} 

\subsection{Properties Of Symmetric Permutation Tree}

Number of nodes at various depths of a $m$ symmetric permutation tree \nameref{notation_SPT} are as follows as shown in Table \ref{table_nodes_depths_spt}.
\begin{enumerate}
\item {
One node with label $i$ where $1\le i\le m$ at depth $1$.}
\item {
 $m - 1$ nodes with label $i$ where $1\le i\le m$ at depth $2$.}
\item {
 $(m - 1)\ast (m - 2)$ nodes with label $i$ where $1\le i\le n$ at depth $3$.}
\item {
 $(m - 1)\ast (m - 2)\ast \ldots \ast (m - j + 1)$ nodes with label $i$ where $1\le i\le m$ at depth $j$ where $1\le j\le m$.}
\item {
 $(m - 1)!$ nodes with label $i$ where $1\le i\le m$ at depth $m$.}
\end{enumerate}

\subsection{Permutation Interpretation of Labeled Regular Bipartite Graph}
\label{sec_per_inter_l_rbg}

Permutation Interpretation of the bi-adjacency matrix of a labeled $r$-regular bipartite graph with $2m$ vertices is obtained in
the following manner.

\begin{enumerate}
\item
We split all the non-zero elements labeled $r$-regular bipartite graph with $2m$ vertices into $r$ sets
such that each set contains exactly one non-zero element or $1$ in exactly one row and column. This decomposition into $r$ sets is clearly not unique.
\item
We associate permutations $p_{1},p_{2},\ldots, p_{r}\in S_{m}$ with each of the $r$ sets.
\item
For each of the $p_{l};1\le l\le r$ sets, if a column $j$ contains a $1$ at row $i$, then the value of the label in the $j^{\mathit{th}}$ location of the permutation is $i$.\\
\noindent for( $u=0;u<r;u$ ++) \{ \\
\hspace{2em}for( $v=0;v<m;v$ ++) \{ \\
\hspace{4em}if a column $v$ contains a $1$ at row $i$ in set $u$, then
the value of the label in the $j^{\mathit{th}}$ location of the
permutation is $i$ in $p_{u}$;\\
\hspace{2em}\}\\
\hspace{0em}\}
\end{enumerate}

\begin{definition} Two Compatible Permutations \\
\label{definition_compper}
Two permutations on a set of $s$ elements represented by $(x_{1}x_{2}\ldots x_{s});x_{p}\neq x_{q}$, such that $\ \forall\ p\neq q;1\le p\le s;1\le q\le s$ for $p,q\in \mathbb{N}$ where $1\le x_{i}\le s$ such that $i\in \mathbb{N};1\le i\le s$ and $(y_{1}\ y_{2}\ \ldots\ y_{s})$ such that $y_{p}\neq y_{q}$
 $\ \forall\ p\neq q$ satisfying $1\le p\le s;1\le q\le s$ for $p,q\in \mathbb{N}$ where 
$1\le y_{i}\le m$ for $i\in \mathbb{N};1\le i\le s$ are compatible if and only if $x_{i}\neq y_{i}\ \forall\ i\in \mathbb{N}$ for $1\le i\le s$.
\end{definition}

We now generalize Definition \ref{definition_compper} for compatible permutations for $r$ permutations in Definition \ref{definition_r_compper}.
\begin{definition} $r$ Compatible Permutations  where $2 \le r \le s$\\
\label{definition_r_compper}
A set of $r$ permutations on a set of $s$ elements represented by 
$(x_{i,1}\ x_{i,2}\ \ldots\ x_{i,s});x_{i,p}\neq x_{i,q}$ $\ \forall\ p\neq q;1\le p\le s;1\le q\le s$ for $p,q\in \mathbb{N}$ where $1\le x_{i,\alpha}\le s\ \forall\ 1\le i\le r;1\le \alpha \le s$ for $i,\alpha \in \mathbb{N}$ are compatible if and only if $x_{i,\alpha }\neq x_{j,\alpha }\ \forall\ i\neq j;1\le \alpha \le s;1\le i\le r;1\le j\le r$ for 
$i,j,\alpha \in \mathbb{N}$. 
\end{definition}

\begin{notation} {Notation for Compatible Permutations} \\
 $p_{i}\notin C(I_{m},p_{2},\ldots, p_{i - 1})$ : $p_{i}$ is compatible with permutations $I_{m},p_{2},\ldots, p_{i - 1}$.
\end{notation}

\begin{theorem}
Any set of $r$ compatible permutations of a set of $s$ elements yields a labeled $r$-regular bipartite graph with $2s$ vertices.
\end{theorem}

\begin{proof}
Let us consider a set of $r$ compatible permutations of a set of $s$ elements represented by $(x_{i,1}x_{i,2}\ldots
x_{i,s});x_{i,p}\neq x_{i,q}\ \forall\ p\neq q;$  $p,q\in \mathbb{N};1\le i\le r;1\le p\le s;1\le q\le s$ where $1\le x_{i,j}\le
s\ \forall\ j\in \mathbb{N}$. By definition, we have $x_{i,p}\neq x_{i,q}\ \forall\ p\neq q;p,q\in \mathbb{N};$ $1\le p\le s;1\le q\le s;\ \forall\ i\in \mathbb{N};1\le i\le r$. This guarantees that all
connections made by the algorithm below are distinct, without any repeated connections between any two set of CN and VN nodes, and we
hence obtain a labeled $r$-regular bipartite graph with $2s$ vertices.
{Let us start with $s$ CN nodes each having distinct labels from $CN_{1}, CN_{2},\ldots,  CN_{s}$, and $s$ VN nodes each having distinct labels from 
$VN_{1} ,VN_{2},\ldots, VN_{s}$.} \\
for( $j=1$ ; $j\le r$ ; $j$ ++) \{ \\
\ for( $i=1$ ; $i\le s$ ; $i$ ++) \ \{ \\
\ \ \ Connect node $CN_{i}$ to node $VN_{x_{i, j}}$;\\
\ \}\\
\}
\end{proof}

\begin{theorem}
Any labeled $r$-regular bipartite graph with $2s$ vertices with the two sets of vertices, CN nodes each having distinct labels $CN_{1}, CN_{2},\ldots,  CN_{s}$, and $s$ VN nodes each having distinct labels from $VN_{1} ,VN_{2},\ldots, VN_{s}$, such that the following conditions are true.
\begin{enumerate}
\item All edges are in the labeled $r$-regular bipartite graph between $CN_{i}$ and $VN_{j}$ for some values of $1 \le i \le s$ and $1 \le j \le s$.
\item There does not exist any edge between any two CN nodes.
\item There does not exist any edge between any two VN nodes.
\end{enumerate}
can be represented by a set of $r$ compatible permutations of a set of $s$ elements.
\end{theorem}
\begin{proof}
Since all the edges in the labeled $r$-regular bipartite graph between $CN_{i}$ and $VN_{j}$ for some values of $1 \le i \le s$ and $1 \le j \le s$, this corresponds to criterion for compatible permutations for $r$ permutations in Definition \ref{definition_r_compper}. Hence, there exists a set of $r$ compatible permutations of a set of $s$ elements that represent labeled $r$-regular bipartite graph. Clearly, this decomposition of a labeled $r$-regular bipartite graph into $r$ compatible permutations is not unique.
\end{proof}

Example \ref{example_3_comp_per} shows three compatible permutations and the bi-adjacency matrix of the corresponding trivalent bipartite graph.

\begin{example} \textbf{Example Of $3$ compatible permutations and the bi-adjacency matrix of the corresponding trivalent bipartite graph} \\
\label{example_3_comp_per}

$\begin{gathered}(\begin{matrix}1&2&3&4&5&6&7&8\end{matrix})\\(\begin{matrix}2&3&4&5&6&7&8&1\end{matrix})\\(\begin{matrix}5&6&7&8&1&2&3&4\end{matrix})\end{gathered}$ $\rightarrow \left[\begin{matrix}1&0&0&0&1&0&0&1\\1&1&0&0&0&1&0&0\\0&1&1&0&0&0&1&0\\0&0&1&1&0&0&0&1\\1&0&0&1&1&0&0&0\\0&1&0&0&1&1&0&0\\0&0&1&0&0&1&1&0\\0&0&0&1&0&0&1&1\end{matrix}\right]$

\end{example}

\subsection{Tree structure after first permutation}

If the first chosen permutation is $(x_{1}x_{2}\ldots x_{s})$ and if
the second permutation chosen is $(y_{1}y_{2}\ldots y_{s})$ where 
$x_{i}\neq y_{i};1\le i\le s$.
{Then the number of successors on the path traversed $(y_{1}y_{2}\ldots y_{s})$ at depths $1,2,\ldots, s$ are $s - 1+\delta_{x_{1},y_{1}},s - 2+\delta _{x_{2},y_{2}},s - 3+\delta_{x_{3},y_{3}},\ldots, 1$}.
{In general for depth $i$ the number of successors on the path traversed are $f(i)=m - i +\delta _{x_{i},y_{i}}$ where $\delta
_{x_{i},y_{i}}=1$ if $x_{i}=y_{j}$ for some $j$ satisfying $s\ge i>j\ge 1$}, {$\delta _{x_{i},y_{i}}=0$ if $x_{i}\neq y_{j}\ \forall\ j$ satisfying $s\ge i>j\ge 1$}, {$\delta _{x_{1},y_{1}}=0$ and $\delta _{x_{s},y_{s}}=1$.} \\
$\delta _{x_{i},y_{i}}=1$ if $x_{i}=y_{j}$ for some $j$ satisfying $s\ge i>j\ge 1$, $\delta _{x_{i},y_{i}}=0$ if $x_{i}\neq y_{j}\ \forall\ j$ satisfying $s\ge i>j\ge 1$.

\begin{theorem}
\label{thm_two_compatible_partitions}
Given any two compatible permutations $a,b\in S_{m}$, we can compute the corresponding partition $\beta (a,b)\in P_{2}(m)$ such that.\\ 
Partition $\beta${ : \ } $S_{m} \times U_{m}(\alpha )\to P_{2}(m)$ where $\alpha $ is the first element chosen from
$S_{m}$, and $U_{m}(\alpha )\cup C(\alpha )=S_{m}$, where $C(\alpha )$ is the set of permutations that are not compatible with $\alpha$.
\end{theorem}

\section{Cycles}
\label{sec_cycles}

The key goal of our research is to understand cycles in regular bipartite graphs and hence it is important to analyze cycles arising between two permutations.

\subsection{Criterion for a Cycle Between Two Permutations}

Two compatible permutations $a,b\in S_{m}$ have a cycle with labeled nodes $u$ and $v$ if Label $u$ at depth $i$ for
permutation $a$ at depth $j$ is Label at depth $j$ for permutation $b$ where $j<i$ and if Label $v$ at depth $i$ for permutation 
$b$ at depth $j$ is Label at depth $k$ for permutation $a$ where $k<i$.
\begin{algorithmo}
\textbf{{Algorithm to calculate $\beta (a,b)\in P_{2}(m)$ given compatible permutations $a,b\in S_{m}$}} \\
\label{algo_cycle_analy_par_2per}
\textbf{Input}: $m \in \mathbb{N}$ and $2$ compatible permutations $a, b \in S_{m}$ such that $\{a, b \}$ represents a $2$-regular bipartite graph. \\
\textbf{Output}: Partition $\beta (a,b)\in P_{2}(m)$ between permutations $a$ and $b$. \\
\textbf{Method}: We assume that the labeled $2$-regular bipartite graph with $2m$ vertices into permutations has been decomposed into compatible permutations $a, b \in S_{m}$. This decomposition of a labeled regular bipartite graph into compatible permutations is not unique.\\
Given compatible permutations $a,b\in S_{m}$, we exhaustively enumerate all cycles by graph traversal and obtain $\beta (a,b)\in P_{2}(m)$ as shown in Figure \ref{func_cal_par_bet_2pers}.  $\Psi (\beta (a,b))$ is isomorphic to the labeled $2$-regular bipartite graph represented by compatible permutations $a,b\in S_{m}$.

\bigskip
\begin{figure}[htbp]
\centering
\vspace{-0.2cm}%
\alglanguage{pseudocode}
\begin{algorithmic}[1]
\Procedure{$CalculatePartitionBetweenPermutations$}{$\{a, b \in S_{m}\}, m$}
\State Temp\_List(a) = Labels in $a$ in increasing order of depth
\State $\beta(a, b)$ = NULL
\While{Temp\_List != NULL}
\State $u$ = Next Label from Temp\_List
\State Remove Label $u$  from Temp\_List
\State $v$ = Position of Label $u$ in permutation $b$
\State $cycle\_len = 0$
\State $d=u$
\While{$d! = v$}
\State $c$ = \ Depth of Label $d$ in permutation $a$
\State  Remove Label $u$  from Temp\_List
\State  $d$ = Label at depth $c$ in permutation $b$
\State   $cycle\_len++$
\EndWhile 
 \State  Add Partition Component $cycle\_len$ to $\beta(a, b)$
\EndWhile
\EndProcedure
\Statex
\end{algorithmic}
  \vspace{-0.4cm}%
\caption{Partition $\beta (a,b)\in P_{2}(m)$ Calculation between permutations $a, b \in S_{m}$}
\label{func_cal_par_bet_2pers}
\end{figure}

\end{algorithmo}

\begin{algorithmo}
\label{algo_cycle_analy_par}
\textbf{{Algorithm to analyze cycles arising from partitions associated with constituent permutations in a labeled $r$-Regular Bipartite Graph}} \\
\textbf{Input}: $m, r \in \mathbb{N}$ such that $r < m$ and $r$ compatible permutations $p_{1}, p_{2}, \ldots, p_{r} \in S_{m}$ such that $\{p_{1}, p_{2}, \ldots, p_{r} \}$ represents a $r$-regular bipartite graph. \\
\textbf{Output}: Partitions $\alpha_{i,j}$ between permutations $p_{i}$ and $p_{j}$ where $i\neq j$ for $1\le i\le r$ and $1\le j\le r$. \\
\textbf{Method}: 
We assume that the labeled $r$-regular bipartite graph with $2m$ vertices into permutations has been decomposed into compatible permutations $p_{1}, p_{2}, \ldots, p_{r} \in S_{m}$. This decomposition of a regular bipartite graph into compatible permutations is not unique.
Calculate partitions $\alpha_{i,j}$ corresponding to $p_{i}$ and $p_{j}$ where $i\neq j$ for $1\le i\le r;1\le j\le r$. Clearly, $\alpha_{i,j}=\alpha_{j,i}\ \forall\ i\neq j$ for $1\le i\le r;1\le j\le r$ as per Algorithm \ref{algo_cycle_analy_par_2per}.
\end{algorithmo}

\begin{theorem}
Given a labeled $r$-regular bipartite graph with $2m$ vertices with compatible permutations $p_{1},p_{2},\ldots, p_{r}$ where the following conditions are valid.
\begin{enumerate}
\item {$p_{i}$ is between $p_{i-1}$ and $p_{i+1}$ on a complete $m$ 
symmetric permutation tree \nameref{notation_SPT} for all integer values of $i$ given by 
$2\le i\le r-1$.}
\item {Partitions between between permutations $p_{i-1}$ and $p_{i}$ 
represented by $\alpha_{i-1,i}\in P_{2}(m)$ for all integer values
of $i$ given by $2\le i\le r-1$.}
\end{enumerate}
then, Partitions $\alpha_{i-1,i}\in P_{2}(m)$ are invariant with any of the
following operations on permutations $p_{1},p_{2},\ldots, p_{r}$
\begin{enumerate}
\item {Position $i$ exchanged with position $j$ where $i\neq j;1\le i\le
m;1\le j\le m$ in each of $p_{1},p_{2},\ldots, p_{r}$.}
\item {Position with value $i$ exchanged with position with value $j$ where 
$i\neq j$ for $1\le i\le m;1\le j\le m$ in each of $p_{1},p_{2},\ldots,p_{r}$.}
\end{enumerate}
\end{theorem}

We introduce the following criterion for length $4$ cycles in a labeled $r$-regular bipartite graph.

\begin{note}{\textbf{Criterion for length $4$ cycles}} \\
{If the permutation representation for a labeled $r$-regular bipartite graph with $2m$ vertices is given
by $\{x_{i,j};1\le i\le r;1\le j\le m\}$, the labeled $r$-regular bipartite graph with $2m$ vertices
has a cycle of length $4$ if $\exists\ l_{1},l_{2}$ such that $l_{1}\neq l_{2};1\le l_{1}\le
m;1\le l_{2}\le m$ satisfying 
$x_{j_{1},l_{1}}=x_{j_{3},l_{2}};x_{j_{2},l_{1}}=x_{j_{4},l_{2}}$ for
some values of $j_{1},j_{2};j_{1}\neq j_{2};1\le j_{1}\le r;1\le
j_{2}\le r$ and $j_{3},j_{4};j_{3}\neq j_{4};1\le j_{3}\le r;1\le
j_{4}\le r$.}
\end{note}

\begin{definition} {Cycle Termination Positions}\\
Given two compatible permutations $p_{1}, p_{2} \in S_{m};p_{2}\notin C(p_{1})$, the cycle termination positions
are the depth positions $1<j\le m$ where the cycles terminate. Each of these cycles correspond to the cycles in the
partition between $p_{1}$ and $p_{2}$.
\end{definition}

\begin{note}{\textbf{Generalized Cycle Traversal}} \\
If labels $l_{1}$ and $l_{2}$ occur at the same depth, labels $l_{2}$ and $l_{3}$ occur at the same depth, $\ldots $, and finally labels
$l_{x}$ and $l_{1}$ occur at the same depth, in the permutation representation of a labeled $r$-regular bipartite graph, then there exists a cycle
connecting the labels $l_{1},l_{2},\ldots, l_{x}$.
\end{note}

\subsection{Partitions For A Labeled Regular Bipartite Graph}

We investigate partitions for a labeled $r$-regular bipartite graph with $2m$ vertices.
The number of partitions for a labeled $r$-regular bipartite graph with $2m$ vertices is specified by a set of permutations $ { p_{1}, p_{2},\ldots, p_{r} }$ is $ r * (r - 1)/2 $ by considering partitions corresponding to all combinations of two permutations $p_{i}, p_{j}$ such that $ i\neq j, 1\le i \le r,  1\le j \le r$.

\begin{definition} $\Phi(\beta _{1},\beta _{2},\ldots,\beta _{r - 1})$ \\
$\Phi (\beta _{1},\beta _{2},\ldots, \beta _{r - 1})$ is the set of all labeled $r$-regular bipartite graphs with $2m$ vertices created with compatible permutations 
$\{p_{1}, p_{2}, \ldots, p_{r}\}$ such that $p_{1}=I_{m}$ and the partition between $p_{i}$ and $p_{i+1}$ is $\beta _{i}$ for $1\le i\le r - 1$.
\end{definition}

\subsection{Important facts about $\Phi (\beta _{1}, \beta _{2}, \ldots, \beta _{r - 1})$}
\begin{enumerate}
\item
Every labeled $r$-regular bipartite graph with $2m$ vertices lies in a unique family of $\Phi (\beta _{1}, \beta _{2}, \ldots, \beta _{r - 1})$.
\item
Of the $r * (r - 1)/2 $ partitions associated with a labeled $r$-regular bipartite graph with $2m$ vertices, the family of $\Phi (\beta _{1}, \beta _{2}, \ldots, \beta _{r - 1})$ specifies only $r$ partitions and hence $r *(r -1)/2 - r = r^2/2 - 3 *r/2$ partitions remain unspecified.
\item
However, many labeled $r$-regular bipartite graphs with $2m$ vertices can correspond to the same non-isomorphic graph. 
Labeled $r$-regular bipartite graphs with $2m$ vertices in multiple families $\Phi(\beta _{1}, \beta _{2}, \ldots, \beta _{r - 1})$ can be isomorphic to each other.
\end{enumerate}

$\Phi (\beta _{1},\beta _{2},\ldots,\beta _{r - 1})$ is a very useful construct for graph construction in a practical context.

\subsection{Known Cycles And Additional Cycles}
For a $r$-regular bipartite graph with $2m$ vertices which is a member of $\Phi (\beta _{1}, \beta_{2}, \ldots, \beta _{r - 1})$ where $\beta _{1}, \beta_{2}, \ldots, \beta _{r - 1}\in P_{2}(m)$,
\begin{enumerate}
\item
The known cycles are due to $\beta _{1}, \beta _{2}, \ldots, \beta_{r - 1}\in P_{2}(m)$ where $\beta _{i}$ is the partition
between $p_{i+1}$ and $p_{i}$ for $1\le i\le r - 1$.
\item
The additional cycles due to other two combinations of permutations $p_{j}$ and $p_{k}$ where $j\neq k,j\neq k+1,1\le j\le r;1\le k\le r$.
\item
Additional Cycles caused due to combinations of three or more permutations.

\end{enumerate}

\section{Labeled Graph Enumeration}
\label{sec_labeled_graph_enumeration}

 We deal with enumeration of labeled $r$-regular bipartite graphs with $2m$ vertices in this section.

\begin{definition} {Permutation Representation of a labeled $r$-regular bipartite graph with $2m$ vertices} \\
\label{definition_perrep}
Permutation Representation of a labeled $r$-regular bipartite graph with $2m$ vertices is specified by specifying the following chord indices.
\begin{equation*}
\{x_{i,j};1\le i\le r;1\le j\le m\}
\end{equation*} 
such that $x_{i, j}\neq x_{t, j}$ for $t\neq i;1\le t\le r$ and $x_{i, j}\neq x_{i, s}$ for $s\neq j;1\le s\le m$.
\end{definition}

Hence, $p_{i}=(x_{i,1}..x_{i,m})\in S_{m};1\le i\le r;$ represent compatible $p_{j}\notin C(p_{1},p_{2},\ldots, p_{j-1})$ for $1<j\le r$.

\subsection{Two distinct kinds of permutations operations}

An element $a\in S_{m}$ could be interpreted in the following two distinct ways.
\begin{enumerate}
\item {Permutations on Depth}.
\item {Permutations on Labels}.
\end{enumerate}

\begin{definition} {Permutations on Depth For a labeled $r$-Regular Bipartite Graph with $2m$ vertices} \\
\label{definition_pod}
A Permutation on Depth is a permutation of depths in the permutation representation of a labeled $r$-regular bipartite graph with $2m$ vertices.
\end{definition}

\begin{definition} {Permutations on Labels For a labeled $r$-Regular Bipartite Graph with $2m$ vertices} \\
\label{definition_pol}
A Permutation on Labels is a permutation of labels in the permutation representation of a labeled $r$-regular bipartite graph with $2m$ vertices, located irrespective of the depths that they are located in.
\end{definition}

\begin{theorem}
Given a labeled $r$-regular bipartite graph with $2m$ vertices with compatible permutations \{$p_{1}, p_{2}, \ldots, p_{r}$\},
Isomorphism is also preserved is preserved with the following operations on permutations $p_{1}, p_{2}, \ldots, p_{r}$
\begin{enumerate}
\item Position $i$ exchanged with Position $j$ where $i\neq j;1\le i\le m;1\le j\le m$ in each of $p_{1}, p_{2}, \ldots, p_{r}$.
\item Position with value $i$ exchanged with Position with value $j$ where $i\neq j;1\le i\le m;1\le j\le m$ in each of $p_{1}, p_{2}, \ldots, p_{r}$.
\end{enumerate}
\end{theorem}

\begin{proof} 
Operation $1$ is equivalent to column exchange of a bi-adjacency matrix of the $r$-regular bipartite graph with $2m$ vertices and Operation $2$ is
equivalent to row exchange of a bi-adjacency matrix of the $r$-regular bipartite graph with $2m$ vertices.
\end{proof}

\begin{theorem}
Given the permutation representation of a labeled $r$-regular bipartite graph with $2m$ vertices, isomorphism is preserved with permutations on Depth and
Permutations on Labels.
\end{theorem}

\begin{proof} 
Permutations on Labels is equivalent to column exchange of a bi-adjacency matrix of the $r$-regular bipartite graph with $2m$ vertices and Permutations on
Depth is equivalent to row exchange of a bi-adjacency matrix of the $r$-regular bipartite graph with $2m$ vertices.
\end{proof}

\begin{theorem}
The Automorphism Group of a labeled $r$-regular bipartite graph with $2m$ vertices can be enumerated by enumerating all
distinct labeled $r$-regular bipartite graphs with $2m$ vertices generated by combinations of permutations on depth and permutations on labels on the equivalent
permutation representation of the labeled $r$-regular bipartite graph with $2m$ vertices.
\end{theorem}

\begin{proof} 
This follows directly from the definition of the Automorphism Group of a labeled $r$-regular bipartite graph with $2m$ vertices, and the
fact that all isomorphisms could be generated by row and column operations on a labeled $r$-regular bipartite graph with $2m$ vertices that results in a
distinct labeled $r$-regular bipartite graph with $2m$ vertices which in turn are equivalent to permutations on depth and permutations on labels.
\end{proof}

\subsection{Two Important Enumeration Problems}
\begin{enumerate}
\item {
Enumeration of the Automorphism Group of a labeled $r$-regular bipartite graph with $2m$ vertices.}
\item {
Enumeration of all non-isomorphic $r$-regular bipartite graph with $2m$ vertices in $\Phi (\beta_{1},\beta _{2},\ldots\, \beta_{r - 1})$.}
\end{enumerate}

\subsection{Enumeration Of Labeled Regular Bipartite Graphs}
Enumeration Of Labeled $r$-regular bipartite graphs with $2m$ vertices consists of the following steps.
\begin{enumerate}
\item {
Enumerating unique non-isomorphic instances of $r$-regular bipartite graph with $2m$ vertices which we
refer to as canonical forms.}
\item {
Enumerating the Automorphism Group for each canonical form of $r$-regular bipartite graph with $2m$ vertices.}
\end{enumerate}
\subsection{Partition Set For A Labeled Regular Bipartite Graph}

We define partition set for a labeled $r$-regular bipartite graph with $2m$ vertices with first permutation $p_{1}=I_{m}$, the identity permutation.

\begin{definition} {Partition Set For A Labeled Regular Bipartite Graph} \\
Partition Set of a labeled $r$-regular bipartite graph with $2m$ vertices with $p_{1}=I_{m}$, $\{p_{1},p_{2},\ldots\, p_{r}\};p_{i+1}\notin C(p_{1},p_{2},\ldots\ ,p_{i})$ for $1\le i\le r - 1$ is generated by combinations of permutations on depths and permutations on labels that result in $p_{1}=I_{m}$ and a different value for one or more of $p_{2},\ldots\ ,p_{r}$.
\end{definition}

\subsection{Automorphism Group Of A Labeled Regular Bipartite Graph}
{For each element in the Partition Set of a labeled $r$-regular bipartite graph with $2m$ vertices, we apply all $m!$ Permutation on depths in order to obtain the Automorphism
Group of a labeled $r$-regular bipartite graph.}

\begin{theorem}
Given $\beta _{1}\in P_{2}(m)$ given by $\sum _{i=1}^{y_{i}} q_{1,i}=m$, consider all possible distinct labeled partitions on the set 
$\{1,2,\ldots, m\}$ such that we have a subset with $q_{1,1}$ elements, a subset with $q_{1,2}$ elements, $\ldots $, a subset with 
$q_{1,y_{1}}$ elements. For each of the above subsets, we consider distinct labeled graphs that arise from variations of permutations
within each subset corresponding to $q_{1,i}$ for $1\le i<y_{1}$ is $\frac{(q_{1,i}!)}{q_{1,i}}$. 
\end{theorem}

\begin{proof} 
If a set has $x$ elements, and if we consider all permutations on this set, with all circular permutations removed, this
has $\frac{(x!)}{x}$ distinct elements. We notice that all circular permutations in each subset results in the same labeled graph, and
hence if we remove all elements that are circular permutations, we obtain the formula $\frac{(q_{1,i}!)}{q_{1,i}}=(q_{1,i}-1)!$ for $1\le i\le y_{1}$.
\end{proof} 

\section{Permutation Enumeration Formulae}
\label{sec_permutation_enumeration_formulae}

We present a permutation enumeration formula in Theorem \ref{thm_pet} for number of compatible permutations.

\begin{theorem} {General Permutation Enumeration formula} \\
\label{thm_pet}
Permutation Enumeration Formulae for compatible permutations 
$\{p_{r};\beta_{r - 1}(p_{r},p_{r - 1}),p_{r}\notin C(p_{1}=I_{m}, p_{2}, \ldots, p_{r -1})\}$ is given by  Equation \ref{per_enu_formula}.
\begin{equation} {\label{per_enu_formula}}
f(\beta _{r - 1})= (m - r + 1) \sum_{j,\mathit{distinct}\ q_{r - 1,j}}{\frac{(m- r + 1)!}{(q_{r - 1,j}-1)\ast \prod_{i=1,i\neq j}^{y_{1}}{(q_{r - 1,i})}}} 
\end{equation}
where $\beta_{r - 1}\in P_{2}(m)$ is given by $\sum_{j = 1}^{y_{l}} q_{z, j} = m$ for $1 \le z \le r - 1$.
\end{theorem}
\begin{proof} 
For each of the possible $m - r + 1$ choices for $p_{r}$ at depth $1$ of the $m$ symmetric permutation tree \nameref{notation_SPT}, the number of distinct permutations that satisfy the constraint that the partition with $p_{r - 1}$ is $\beta_{r-1}$ are $\sum_{j,\mathit{distinct}\ q_{r - 1,j}}{\frac{(m- r + 1)!}{(q_{r - 1,j}-1)\ast \prod_{i=1,i\neq j}^{y_{1}}{(q_{r - 1,i})}}}$, $\beta _{r - 1}\in P_{2}(m)$ since we get different permutations only for for each distinct $q_{r - 1, j}$ and hence the result in Equation \ref{per_enu_formula} follows.
\end{proof}

\begin{corollary} 
Permutation Enumeration Formulae for compatible permutations 
$\{p_{2};\beta _{1}(p_{2},p_{1}),p_{2}\notin C(p_{1}=I_{m})\}$ is given by Equation \ref{per_enu_formula_compatible_per}.
\begin{equation}  {\label{per_enu_formula_compatible_per}}
f(\beta _{1})= (m - 1) *\sum
_{j,\mathit{distinct}q_{1,j}}{\frac{(m -1)!}{(q_{1,j}-1)\ast \prod
_{i=1,i\neq j}^{y_{1}}{(q_{1,i})}}} 
\end{equation}
where $\beta _{1}\in P_{2}(m)$ is given by $\sum_{j = 1}^{y_{1}} q_{1, j} = m$.
\end{corollary} 

\begin{corollary} 
\label{cor_per_enu}
Permutation Enumeration Formulae for compatible permutations\\ 
$\{p_{2};\beta _{1}(p_{2},p_{1}) = (m), p_{2}\notin C(p_{1})\}$ is given by Equation \ref{per_enu_formula_full_cycle}.
\begin{equation} {\label{per_enu_formula_full_cycle}}
f\{\beta _{1} = (m)\}= {(m - 1)!}
\end{equation}
where $\beta _{1} = (m) \in P_{2}(m)$.
\end{corollary}

\section{Towards Exhaustive Enumeration}
\label{sec_towards_exhaus_enu}

We introduce the following thought experiment in order to visualize all labeled $r$-regular bipartite graphs with $2m$ vertices that can be created by choosing permutations on a $m$ symmetric permutation tree, \nameref{notation_SPT}.
\subsection{Thought Experiment}
\begin{enumerate}
\item {
Start with a set of $r$ compatible partitions $q_{1}, q_{2}, \ldots, q_{r}$ where without loss of generality, $q_{2},\ldots, q_{r}$
occur in the order of traversal of all leaf nodes of a complete $m$ 
symmetric permutation tree, \nameref{notation_SPT} from first element $I_{m}$ until the last
element.}
\item {
Vary $q_{r}$ for all possible compatible leaf nodes on the complete 
$m$ symmetric permutation tree, \nameref{notation_SPT}.}
\item {
For each of the above $q_{r}$ above, we vary $q_{r-1}$ for all
possible compatible leaf nodes on the complete $m$ symmetric
permutation tree.}
\item {
This process continues until for each of the possible $q_{3}$, we vary $q_{2}$ for all possible compatible leaf nodes on the complete $m$ 
symmetric permutation tree, \nameref{notation_SPT}.}
\end{enumerate}

\subsection{Exhaustive Enumeration Of Leaf Nodes Of Symmetric Permutation Tree}

We describe \nameref{notation_ESPT}, Algorithm \ref{algo_enu_exh_spt} for enumerating all leaf nodes of \nameref{notation_ESPT}.

\begin{notation}
\subsubsection{ESPT$(m)$}
\label{notation_ESPT}
We use the notation \nameref{notation_ESPT} to refer to Algorithm \ref{algo_enu_exh_spt} for Exhaustive Enumeration Of Leaf Nodes Of $m$ Symmetric permutation tree, \nameref{notation_SPT}.
\end{notation}

\begin{algorithmo} {\label{algo_enu_exh_spt}} \nameref{notation_ESPT} \\ \textbf{{Exhaustive Enumeration Of Leaf Nodes Of Symmetric Permutation Tree} \\
\textbf{Input}: $m \in \mathbb{N}$}\\
\textbf{Output}: All distinct permutations $p \in S_{m}$ represented by $\{x_{1}\ x_{2}\ \ldots\ x_{m}\}$ \\
\textbf{Method}: We enumerate a $m$ Symmetric Permutation Tree, \nameref{notation_SPT} by invoking\\ $P(\{NULL, 0\}, \{(1, 2, \ldots, m), m\})$ in $ESPT(m)$ as shown in Figure \ref{func_exhaust_enu_spt} and Figure \ref{func_enu_spt}. Here $(1, 2, \ldots, m)$ is the initial permutation for the enumeration algorithm. However, without loss of generality, we could have chosen any other permutation $q \in S_{m}$. The total number of leaf nodes of a $m$ symmetric permutation tree, \nameref{notation_SPT} is given by the recurrence Equation \ref{eq_recur_spt} and Equation \ref{eq_recur_spt_one}.
\begin{equation}
\label{eq_recur_spt}
T(m) = m \ast T(m - 1)\  \mathit{for}\ m > 1
\end{equation}
\begin{equation}
\label{eq_recur_spt_one}
T(1) = 1
\end{equation}

\bigskip
\begin{figure}[htbp]
\centering
\vspace{-0.2cm}%
\alglanguage{pseudocode}
\begin{algorithmic}[1]
\Procedure{$ESPT$}{$m$} \\
	\  \ \  \ {$P$}{$(\{NULL, 0\}, \{(1, 2, \ldots, m),m\})$} 
\EndProcedure
\Statex
\end{algorithmic}
  \vspace{-0.4cm}%
\caption{Exhaustive Enumeration Of Symmetric Permutation Tree}
\label{func_enu_spt}
\end{figure}

\bigskip
\begin{figure}[htbp]
\centering
\vspace{-0.2cm}%
\alglanguage{pseudocode}
\begin{algorithmic}[1]
                \Procedure{$P$}{$\{(y_{1}, \ldots, y_{r}), r\}, \{(x_{1},x_{2},\ldots, x_{n}),n\}$}
\If {${n}$ == 0}
     \State \textbf{return $(y_{1}\ y_{2}\ \ldots\ y_{r})$}
\EndIf
\For{$j = 1 \to n$} 
	\State {$P$}{$(\{(y_{1}, \ldots, y_{r}, x_{j}\}, r + 1), \{(x_{1},x_{2},\ldots, x_{j - 1}, x_{j + 1}, \dots, x_{n}),n - 1\})$}
	\EndFor
\EndProcedure
\Statex
\end{algorithmic}
 \vspace{-0.4cm}%
\caption{Function For Exhaustive Enumeration Of Leaf Nodes}
\label{func_exhaust_enu_spt}
\end{figure}
\end{algorithmo}

\begin{theorem} \nameref{notation_ESPT}, Algorithm \ref{algo_enu_exh_spt} enumerates all $\{m!\}$ elements of $S_{m}$.
\end{theorem}
\begin{proof} We split the proof into three parts.
\begin{enumerate}
\item At depth $1$, a $m$ symmetric permutation tree, \nameref{notation_SPT} has $n$ nodes each corresponding to the $m$ invocations of $P(\{j, 1\}, \{(1, 2, \ldots, j, j + 1, \ldots  m), m - 1\})$ generated by invoking $P(\{NULL, 0\}, \{(1, 2, \ldots, m), m\})$. Similarly, there are $m - 1$ successor nodes at depth $2$ for each node at depth $1$ and finally at depth $m$, there is $1$ successor node for each node at depth $m - 1$. This proves that the number of permutations generated at depth $m$ is $m \ast (m - 1) \ast \ldots = \{m!\}$ which is the solution for Equation \ref{eq_recur_spt} and Equation \ref{eq_recur_spt_one}. 
\item In addition, we notice that during each invocation of \\ $P(\{(y_{1}, \ldots, y_{r}), r\}, \{(x_{1},x_{2},\ldots, x_{m - r}),m - r\})$, we have $m - r$ invocations of \\$P(\{(y_{1}, \ldots, y_{r}, x_{j}\}, r + 1), \{(x_{1},x_{2},\ldots, x_{j - 1}, x_{j + 1}, \dots, x_{m - r}),m - r - 1\})$, each of these corresponds to node $x_{j}$ at depth $r$ of the symmetric permutation tree.
\item We can thus establish a one-one onto map between the leaf nodes of a $m$ symmetric permutation tree, \nameref{notation_SPT} and the $(y_{1}\ y_{2}\ \ldots\ y_{m})$ returned at depth $m$ have all distinct elements which are some permutation of the initial permutation $(1\ 2\ \ldots\ m)$.
\end{enumerate}
Hence, Algorithm \ref{algo_enu_exh_spt} enumerates all $\{m!\}$ elements of $S_{m}$.
\end{proof}

\begin{corollary} Algorithm \ref{algo_enu_exh_spt} is in $EXPTIME$.
\end{corollary}

\subsection{Maximum Attainable Girth For Degree 2}

The maximum attainable girth of a $2$-regular bipartite graph with $2m$ vertices is stated in Theorem \ref{thm_max_girth_deg2}.
\begin{theorem} 
\label{thm_max_girth_deg2}
{The maximum possible girth of a $2$-regular bipartite graph with $2m$ vertices is $2m$.}
\end{theorem}
\begin{proof} 
This directly follows when we consider that every $2$-regular bipartite graph with $2m$ vertices can be mapped to $\Psi (\beta )$ where $\beta \in P_{2}(m)$. It is clear that girth of a $2$-regular bipartite graph with $2m$ vertices is $\{2\mathit{min}(q_{i})\}$ where $1\le i\le y$ and $\sum _{i=1}^{y}q_{i}=m$ represents $\beta \in P_{2}(m)$. Hence, it follows that the maximum possible girth of a $2$-regular bipartite graph with $2m$ vertices is $2m$.
\end{proof} 

The cycles of a $2$-regular bipartite are known when the number of vertices $2m$ and the partition $\beta \in P_{2}(m)$ is known.

\begin{theorem} {Known Cycle Theorem for a Regular Bipartite Graph of degree $2$}\\
\label{thm_known_cycles}
The cycle lengths of a $2$-regular bipartite graph with $2m$ vertices that is isomorphic to $\Psi (\beta )$ for some $\beta \in P_{2}(m)$ given by $\sum _{i=1}^{y}q_{i}=m$ are $\{2q_{i}\}$ where $1\le i\le y$.
\end{theorem}
\begin{proof} 
A $2$-regular bipartite graph with $2m$ vertices that is isomorphic to $\Psi (\beta )$ has no other cycles other than that of $\beta \in P_{2}(m)$ given by $\sum _{i=1}^{y}q_{i}=m$. The cycle length for a partition component $q_{i}$ is $2q_{i}$. Hence, it follows that the cycle lengths are $\{2q_{i}\}$ where $1\le i\le y$.
\end{proof}

\section {Conclusion}
\label{sec_rbg_chap_summ}
This paper describes a general algorithm for all non-isomorphic $2$-regular bipartite graphs with $2m$ vertices and a mathematical proof has been provided for its completeness. Several results for $E(m,r)$ have been proved. An abstraction of Symmetric Permutation Tree in order to visualize a labeled $r$-regular bipartite graph and enumerate its automorphism group has been introduced. An algorithm to generate the partition associated with two compatible permutations has been introduced. The relationship between Automorphism Group and permutation enumeration problem has used to derive formulae for compatible permutations.



\end{document}